\documentclass{iopart}

\usepackage{graphicx}  
\usepackage{latexsym}
\usepackage{amssymb}

\jl{6}        
\eqnobysec    

\def\beq{\begin{equation}}
\def\eeq{\end{equation}}
 
\def\eqref#1{(\ref{#1})}

\def\rightcontract{\mathop{\hbox{\vrule width0.5pt height6pt%
  \vrule height0.5pt width6pt}}}

\begin{document}

\title[On the local isometric embedding of trapped surfaces]
{On the local isometric embedding of trapped surfaces into 
 three-dimensional Riemannian manifolds}

\author{
Donato Bini${}^{1,2}$
and 
Giampiero  Esposito$^{2}$
}

\address{
  ${}^1$\
  Istituto per le Applicazioni del Calcolo ``M. Picone,'' CNR, Via dei Taurini 19, I-00185, Rome, Italy
}

\address{
  ${}^2$
  INFN, Sezione di Napoli, Complesso Universitario di Monte S. Angelo,\\
Via Cintia Edificio 6, 80126 Naples, Italy 
}

 \date{\today}

\begin{abstract}
We study trapped surfaces from the point of view of local isometric embedding
into three-dimensional Riemannian manifolds. When a two-surface is embedded into three-dimensional
Euclidean space, the problem of finding all surfaces applicable upon it gives rise to a
non-linear partial differential equation of the Monge-Amp\`ere type, first discovered by Darboux, 
and later reformulated by Weingarten. Even today, this problem remains very difficult,
despite some remarkable results. We find an original way of
generalizing the Darboux technique, which leads to a coupled set of $6$ non-linear partial 
differential equations. For the $3$-manifolds occurring in 
Friedmann-(Lemaitre)-Robertson-Walker cosmologies, we show that the local isometric embedding
of trapped surfaces into them can be proved by solving just one non-linear equation.
Such an equation is here solved for the three kinds of Friedmann model associated with positive, 
zero, negative curvature of spatial sections, respectively.
\end{abstract}

\maketitle

\section{Introduction}
\setcounter{equation}{0}

Since Roger Penrose wrote his Adams Prize essay paper on space-time geometry \cite{PE1965},
the concept of trapped surface has received careful consideration in the modern literature on
general relativity, and it was only in the year 2009 that rigorous results on the origin of
black holes through the formation of trapped surfaces were collected in a  monograph
by Christodoulou \cite{BH2009}, further improved, later, by Klainerman and Rodnianski
\cite{KR2012}. The definition of trapped surfaces regards them as $2$-dimensional
Riemannian manifolds whose two null normals have expansion (also
denoted as null mean curvature) which is always negative, so that neighbouring light
rays, normal to the surface, must move towards one another \cite{Irina,Yvonne,BE2017,Sherif}.
The boundary of a connected component of the trapped region that contains the trapped surface
is an apparent horizon. A weaker concept is also of interest, i.e. a marginally outer trapped
surface is a closed $(d-2)$-dimensional Riemannian mamifold in a $d$-dimensional
space-time, whose outer null normals have vanishing expansion. 

The aim of our paper is to fully exploit the classical differential geometry of surfaces
developed by Gauss \cite{GA}, Darboux \cite{DA1,DA2,DA3,DA4}, Weingarten \cite{WE1896},
Ricci-Curbastro \cite{RI1898} and Bianchi \cite{BI1,BI2,BI3,BI4}, and apply it to an investigation
of the broadest family of trapped surfaces, without making any restriction to problems with
spherical symmetry or yet other symmetry. For this purpose, section 
$2$ studies the most general form of 
metric on a trapped surface. This effort is important because, although everybody agrees on the 
general definition, according to which the two null mean curvatures are both negative
\cite{SE2013,BE2013,MN2015}, only a few explicit examples have been studied so far, to the best
of our knowledge. Section $3$ describes first-order differential properties of null
congruences with the associated Newman-Penrose formalism, and an explicit example of trapped
surface in Schwarzschild spacetime is studied.
In section $4$ we study the problem of local isometric embedding of surfaces into
Riemannian 3-manifolds, obtaining eventually a set  
of coupled non-linear equations which generalize the Darboux technique for the
isometric embedding of a surface into three-dimensional Euclidean space.
Section $5$ restricts attention to a case tractable by a single non-linear equation,
i.e. the local isometric embedding of a surface into the 3-dimensional spatial
sections of Friedmann cosmology. Explicit solutions of this equation are obtained in
section $6$ for the 3 families of Friedmann cosmologies, when the unknown function 
depends only on the angular variable $\theta$. Concluding remarks are presented in section 
$7$, while $3$ appendices summarize some technical properties. 

\section{General form of the metric on a trapped surface}
\setcounter{equation}{0}
 
The concept of trapped surface involves the interplay of Riemannian and 
pseudo-Riemannian geometry \cite{LC1926,ON1983} 
in dimension $2,3$ and $4$. The space-time manifold is here denoted by 
$(V,g)$, where $V$ is connected, four-dimensional, Hausdorff and $C^{\infty}$. The space-time
metric $g$ is Lorentzian, with signature $(-,+,+,+)$. Strictly, we are dealing with an
equivalence class of pairs $(V,g)$, where any two metrics
$g$ and ${\hat g}$ are equivalent if there exists a
diffeomorphism mapping $g$ into ${\hat g}$ \cite{HE1973}. 
We will use the notation for the inner product\footnote{We will include $g$ in the definition 
of inner product, i.e. $<A,B>_g$, only when necessary, i.e. when it involves vectors of 
$2$- and $3$-dimensional sub-spaces.} associated with $g$: 
$<A,B> \equiv (g^{-1})^{\alpha\beta}A_\alpha B_\beta$.
The Einstein convention of summation over repeated Greek indices is here used,
but in cases where a mixture of tensor formulae in $2$, $3$ and $4$ dimensions occurs,
we shall restore the explicit summations.

A globally defined timelike vector field $X: M \rightarrow TM$ (so that $X(p)\in T_p (M))$, $p\in M$ is 
also taken to exist, that defines a time orientation. A timelike or null
tangent vector $u \in T_{p}(M)$ is said to be future-directed if $g(X(p),u)<0$, or past-directed if
$g(X(p),u)>0$ \cite{HE1973}.

Furthermore, we consider a {\it family} of spacelike $3$-manifolds 
$(\Sigma_3 ,g_{3})$ with unit {\it timelike} normal vector field $n$,
and a {\it family} of Riemannian $2$-manifolds $(\Sigma_2,H)$ with unit {\it spacelike} normal $\hat\nu$. 
The proper time along the congruences of world lines with unit tangent $n$, say $\tau_n$, 
can be used to parameterize the hypersurfaces of the family $\Sigma_3\equiv (\Sigma_3^{(\tau_n)})$. 
An additional parameter $\ell_{\hat \nu}$ (curvilinear abscissa along the world lines of the 
congruence $\hat\nu$) is then necessary to parameterize $\Sigma_2\equiv (\Sigma_2^{(\tau_n, \ell_{\hat \nu})})$.

One can form null vector fields in
space-time out of linear combinations $l^{\pm}$ of the normal vectors $n$ and $\hat\nu$, 
proportional to $(n \pm \hat \nu)$.
Moreover, since $n$ can be considered as the $4$-velocity field of a family of test observes filling some 
space-time region (i.e., the region where their causality condition is preserved) we have that the 
spatial (with respect to $n$) vectors $\pm \hat \nu\equiv \hat \nu (l^\pm, n)$ represent the directions of the 
relative velocities (the magnitude is always $1$).

We can write then 
\begin{equation}
l^{+} \equiv E(l^+,n)(n+\hat\nu), \qquad l^{-} \equiv E(l^-,n)(n-\hat\nu)\,,
\label{(2.1)}
\end{equation}
where we have introduced the relative energy factors $E(l^\pm,n)$ (which are not constant in general).
We will require that both $l^+$ and $l^-$ be also {\it geodesic} affinely parameterized, $\nabla_{l^\pm}l^\pm=0$.
No further assumptions are considered on $n$ and $\hat \nu$ below.

The normals to the family of hypersurfaces $\Sigma_3$ and $\Sigma_2$ can be therefore re-expressed in the form
\begin{equation}
\fl\quad
n={1 \over 2 }\left(\frac{l^{+}}{E(l^+,n)}+\frac{l^{-}}{E(l^-,n)}\right)\,,\qquad
\hat \nu={1 \over 2 }\left(\frac{l^{+}}{E(l^+,n)}-\frac{l^{-}}{E(l^-,n)}\right)\,, 
\label{(2.2)}
\end{equation}
and the space-time tensor field 
\begin{equation}
h \equiv g+ n \otimes n - \hat\nu \otimes \hat\nu
\label{(2.3)}
\end{equation}
reads eventually as
\begin{equation}
h=g+ \frac{1}{2 E(l^+,n)E(l^-,n)}\left(l^{+} \otimes l^{-}+l^{-} \otimes l^{+}\right),
\label{(2.4)}
\end{equation}
because the coefficients of $l^{+} \otimes l^{+}$ and $l^{-} \otimes l^{-}$
vanish exactly. The condition for $l^{+}$ and $l^{-}$  to be null vectors  implies that
\begin{equation}
\frac{1}{E(l^+,n)}\langle l^{\pm},l^{\pm} \rangle
= \langle n,n \rangle + \langle \hat\nu,\hat\nu \rangle 
\pm 2 \langle n,\hat\nu \rangle=0,
\label{(2.5)}
\end{equation}
Recalling that $n$ (timelike) and $\hat\nu$ (spacelike)  are unit vectors, 
$\langle n,n \rangle=-1$ and $\langle \hat\nu,\hat\nu \rangle=1$, 
the above conditions (\ref{(2.5)})  
imply
$\langle n,\hat\nu \rangle=0$ 
and hence
\begin{equation}
\langle l^{+},l^{-} \rangle=-\frac{2}{E(l^+,n)E(l^-,n)}\,.
\label{(2.6)}
\end{equation}
 
The tensor field (\ref{(2.4)}) is a space-time projector and satisfies the property of 
squaring to itself, that is
\begin{equation}
\label{(2.7)}
h^2=h \rightcontract h = h,
\end{equation}
as can be easily proven with a straightforward, direct calculation. In the above expression the symbol 
$\rightcontract$ denotes right contraction among tensors, e.g., 
$(A\rightcontract B)^\alpha{}_\beta=A^\alpha{}_\mu B^\mu{}_\beta$ for any 2-tensors $A$ and $B$;
$h$ projects orthogonally onto both $n$ and $\hat \nu$.

The two null mean curvatures are defined as
\begin{equation}  
\chi^{+} \equiv 
(h^{-1})^{\alpha \beta}\nabla_{\alpha}l^+_{\beta},\qquad
\chi^{-} \equiv 
(h^{-1})^{\alpha \beta}\nabla_{\alpha}l^-_{\beta}.
\label{(2.8)}
\end{equation}
They can be computed for all surfaces within the considered family, and then restricted to a single one 
with a specific choice of the parameters.

The projector $h$ has covariant components 
\begin{equation}
h_{\mu \nu}=g_{\mu \nu}+ \frac{1}{2 E(l^+,n)E(l^-,n)} \left(l^+_{\mu}l^-_{\nu}+l^-_{\mu}l^+_{\nu}\right), 
\label{(2.9)}
\end{equation}
and induces the metric $H$ on the surface we are studying, which is a Riemannian 
$2$-manifold $(\Sigma_2,H)$ \footnote{
In other words, if $p \in \Sigma \subset V$,
then the Riemannian nature of $(\Sigma_2,H)$ implies that the tangent space to $V$ at $p$
can be split as the direct sum
\begin{equation}
T_{p}V=T_{p}\Sigma \oplus (T_{p}\Sigma)^{\perp},
\label{(2.10)}
\end{equation}
and there exists a unique lift of $h$ from the space of symmetric rank-$2$ tensors
\begin{equation}
{\rm Sym}_{2}(T_{p}\Sigma_2) \equiv \left \{ \sum_{a,b=1}^{2}T_{ab}{\rm d}x^{a} \otimes {\rm d}x^{b} 
\right \} ,
\label{(2.11)}
\end{equation}
to the space of symmetric rank-$2$ tensors
\begin{equation}
{\rm Sym}_{2}(T_{p}V) \equiv \left \{ \sum_{\mu,\nu=1}^{4}\tau_{\mu \nu}{\rm d}x^{\mu} 
\otimes {\rm d}x^{\nu} 
\right \} .
\label{(2.12)}
\end{equation}
}.
In order to avoid confusion, we denote by $H_{ab}$ the components of the Riemannian metric
on $\Sigma_2$, with $a,b=1,2$, and by $h_{\mu \nu}$ the components (2.9) of the projector $h$,
with $\mu,\nu=1,...,4$.

The surface $(\Sigma_2,H)$ is said to be trapped if the null mean curvatures defined in (2.8) 
are both negative, i.e. $\chi^{+}<0$ and $\chi^{-}<0$. 
In Eqs. (2.8), the traces $\chi^{+}$ and $\chi^{-}$ obtained from $(h^{-1})^{\alpha \beta}$
reduce to the traces obtained from the contravariant metric $(g^{-1})^{\alpha \beta}$, i.e.
\begin{equation}
\chi^{+}=(g^{-1})^{\alpha \beta}\nabla_{\alpha}l^+_{\beta}.
\label{(2.13)}
\end{equation}
We postpone the proof of this relation in Appendix A.
The limiting case of a surface where $\chi^\pm=0$ corresponds to an apparent horizon.

\section{First-order differential properties of  null congruences and Newman-Penrose formalism}
\setcounter{equation}{0}

For any null geodesic congruence of world lines affinely parameterized, say $l$ (later to be either $l^+$ 
or $l^-$) one defines the following invariant quantities (optical scalars, not depending on the choice of 
the frame): expansion $\Theta(l)$, vorticity or twist $\omega(l)$  and shear $|\sigma(l)|$ such that
\begin{equation}
\Theta(l)=\frac12 \nabla_\alpha l^\alpha ,\qquad \omega^{2}(l)
=\frac12 \nabla_{[\alpha} l_{\beta]}\nabla^{\alpha} l^\beta
\label{(3.1)}
\end{equation}
and
\begin{equation}
|\sigma(l)|^2+\Theta^{2}(l)=\frac12 \nabla_{(\alpha} l_{\beta)}\nabla^{\alpha} l^{\beta} . 
\label{(3.2)}
\end{equation}
The tensor $\nabla_{\alpha} l_\beta\equiv F_{\alpha\beta}$ obeys 
\begin{equation}
{}^*F_{\alpha\beta} F^{\alpha\beta}=0,
\label{(3.3)}
\end{equation}
(the rank of the matrix $\nabla^{\alpha} l^\beta$ is maximally two); this and the identity 
$\eta^{\alpha \beta \gamma\delta }\nabla_{\delta} \nabla_{\alpha}l_{\gamma} = 0
=\eta^{\alpha \beta \gamma\delta }\nabla_{\delta}F_{ \alpha\gamma}$ 
implies
\begin{equation}
\nabla_{\alpha} (\omega(l)\,  l^\alpha)= \omega(l)_{,\alpha}l^\alpha + 2\Theta(l)\omega(l)=0\,.
\label{(3.4)}
\end{equation}
The (complex) combination 
\begin{equation}
\rho(l) = -(\Theta(l)+i \omega(l))
\label{(3.5)}
\end{equation}
is also one of the $12$ Newman-Penrose \cite{NP1962} spin coefficients.
For a more complete discussion about the formation of trapped surfaces in Schwarzschild 
space-time we refer to Ref. \cite{Wald:1991zz}.
In what follows we limit ourselves to showing 
the (known) results for the expansion of null congruences in the Schwarzschild 
and Schwarzschild-de Sitter metrics.

Let us consider the $2$-spheres $t=$const and $r=$const in the 
Schwarzschild space-time, with standard metric
\begin{equation}
\fl\qquad
g= -N^2 {\rm d}t \otimes {\rm d}t +N^{-2} {\rm d}r \otimes {\rm d}r 
+r^{2} ({\rm d}\theta \otimes {\rm d}\theta
+\sin^2\theta {\rm d}\phi \otimes {\rm d}\phi),
\label{(3.6)}
\end{equation}
where the lapse function $N$ is given by
\begin{equation}
N=\sqrt{1-\frac{2M}{r}}.
\label{(3.7)}
\end{equation}
In this case the unit normals to the $t=$const. and $r=$const. hypersurfaces are
\begin{equation}
n=\frac{1}{N}{\partial \over \partial t}, \qquad \hat \nu= N{\partial \over \partial r},
\label{(3.8)}
\end{equation}
respectively, so that the two null vector fields naturally associated with them are
\begin{equation}
l^\pm =\frac{E(l^\pm ,n)}{N}\left({\partial \over \partial t} 
\pm  N^2 {\partial \over \partial r}  \right).
\label{(3.9)}
\end{equation}
Upon choosing the energy factor $E(l^\pm ,n)$ as
\begin{equation}
E(l^\pm ,n)=\frac{{\mathcal E}}{N},\qquad {\mathcal E}={\rm const.}
\label{(3.10)}
\end{equation}
both $l^\pm$ turn out to be geodesic (affinely parametrized)
\begin{equation}
l^\pm =\frac{{\mathcal E}}{N^2}\left({\partial \over \partial t} 
\pm  N^2 {\partial \over \partial r}  \right)
=\frac{{\mathcal E}}{N^2}{\partial \over \partial t} 
\pm  {\mathcal E}{\partial \over \partial r}.
\label{(3.11)}
\end{equation}
Moreover
\begin{equation}
\chi^\pm = \nabla_\alpha l^\pm{}^\alpha = \pm 2\frac{{\mathcal E}}{r}\,,
\label{(3.12)}
\end{equation}
or
\begin{equation}
\Theta(l^\pm)=\pm \frac{{\mathcal E}}{r}\,.
\label{(3.13)}
\end{equation}

The situation is similar on passing to the case of Schwarzschild-de Sitter spacetime, 
with the same metric element as \eqref{(3.6)} but with lapse function
\begin{equation}
N=\sqrt{1-\frac{2M}{r}-\Lambda r^2},
\label{(3.14)}
\end{equation}
with $\Lambda$ the cosmological constant.
This solution is such that
\begin{equation}
R_{\alpha\beta}=3\Lambda g_{\alpha\beta}.
\label{(3.15)}
\end{equation}
The null geodesic vector fields read as in Eq. \eqref{(3.11)} 
(with $N$ given by \eqref{(3.14)} and lead to the same Eq. \eqref{(3.12)}.

\subsection{Friedmann-Lemaitre--Robertson-Walker spacetime}
Let us consider the case of a Friedmann-(Lemaitre)-Robertson-Walker (FLRW) spacetime, 
with metric (strictly, Lemaitre did not consider the spatially flat case)
\begin{equation}
ds^2=-dt^2 +a^{2}(t)\left[\frac{dr^2}{(1-kr^{2})}+r^2 (d\theta^2+\sin^2\theta d\phi^2)  \right]
\label{(3.16)}
\end{equation}
where the coordinates $(t,r,\theta,\phi)$ are referred to as \lq\lq reduced-circumference 
polar coordinates." Here $a(t)$ is the (dimensionless) scale factor whereas $k=-1,0,1$ corresponds 
to a three-geometry with negative, zero, and positive curvature, respectively.
Note that $k$ has units of length$^{-2}$ and  $r$ has units of length. Moreover, a) $k$ results in 
the Gaussian curvature of the space at the time when $a(t) = 1$; b) $r$ is sometimes called the 
\lq\lq reduced circumference" because it is equal to the measured circumference of a circle (at that 
value of $r$), centered at the origin, divided by $2\pi$, like the $r$ of Schwarzschild coordinates.

The metric \eqref{(3.16)} is a solution to Einstein's field equations $G_{\mu \nu }+\Lambda g_{\mu \nu }
= 8\pi T_{\mu \nu } $  giving the Friedman  equations when the energy-momentum tensor is  
assumed to correspond to a perfect fluid, with  isotropic and homogeneous pressure and energy 
density. The resulting equations are
\begin{eqnarray}
&&\left(\frac{\dot a}{a}\right)^{2} + \frac{k }{a^2} - \frac{\Lambda }{3} = \frac{8\pi}{3}\rho\nonumber\\
&& 2{\frac {\ddot {a}}{a}}+\left({\frac {\dot {a}}{a}}\right)^{2}+{\frac {k }{a^{2}}}-\Lambda =- 8\pi p,  
\label{(3.17)}
\end{eqnarray}
or, equivalently
\begin{eqnarray}
{\dot \rho} &=& - 3 \frac{\dot a}{a} (\rho+p)\nonumber\\
\frac {\ddot {a}}{a}&=&- \frac {4\pi }{3}  (\rho +3p)+ \frac {\Lambda }{3} ,
\label{(3.18)}
\end{eqnarray}
with $k$, the spatial curvature index, serving as an integration constant for the first equation.
The radial (ingoing, outgoing) null geodesics are given by (see Section $6$ for the
application of these formulae)
\begin{equation}
l^\pm =\frac{1}{a(t)}\left({\partial \over \partial t} \pm 
\frac{\sqrt{1-kr^2}}{a(t)} {\partial \over \partial r} \right),
\label{(3.19)}
\end{equation}
i.e., along the null orbits $r(t)$ varies according to
\begin{equation}
\frac{dr}{dt}\equiv \dot r =\pm \frac{\sqrt{1-kr^2}}{a(t)}.
\label{(3.20)}
\end{equation}
In this case $\chi^\pm$ depends on $t$ (and $k$), i.e.,
\begin{equation}
\fl\quad
\chi^\pm = \nabla_\alpha l^\pm{}^\alpha =\frac{2}{a^2  }\left[ \dot a  \pm 
\frac{\sqrt{1-k r^2}}{r  } \right]=
\frac{2}{a  }\left[-\frac13 \frac{\dot \rho}{(p+\rho)} \pm \frac{\sqrt{1-k r^2}}{r a} \right].
\label{(3.21)}
\end{equation}
Equivalently, along the null orbits we have then 
\begin{equation}
\chi^\pm =\frac{2}{a^2  }\left( \dot a + \frac{a \dot r}{r  } \right)=\frac{2}{a }\left( 
\frac{\dot a}{a}  +\frac{\dot r}{r  } \right)
=\frac{2}{a }\frac{d}{dt} \ln (ar) .
\label{(3.22)}
\end{equation}
Note also that from the first of Eqs. \eqref{(3.17)} one can re-express $\dot a/a$ as
\begin{equation}
\frac{\dot a}{a}=\pm \frac{1}{a}\sqrt{\frac{8\pi\rho a^2 }{3}-k+\frac{\Lambda a^2}{3}}.
\label{(3.23)}
\end{equation}
Let us limit our considerations to the case ${\dot a}/{a}>0$ corresponding to an expanding Universe.
We find
\begin{equation}
\chi^\pm =\frac{2}{a^2  }\left[\sqrt{\frac{8\pi\rho a^2 }{3}-k+\frac{\Lambda a^2}{3}}  
\pm \frac{\sqrt{1-k r^2}}{r  } \right]. 
\label{(3.24)}
\end{equation}
In the absence of cosmological constant, the above relation becomes
\begin{equation}
\chi^\pm =\frac{2}{a^2  }\left[\sqrt{\frac{8\pi\rho a^2 }{3}-k }  
\pm \frac{\sqrt{1-k r^2}}{r  } \right]. 
\label{(3.25)}
\end{equation}
and coincides (up to an overall factor of $1/a$) with the result in \cite{SHthesis} 
[See pag. 13 therein, recalling that in Ref. \cite{SHthesis} the energy density is denoted by  
$\mu$ in place of our $\rho$ and units are such that  $8\pi=1$ there.]. In the cases 
$k=-1,0$ Eq. \eqref{(3.25)} shows that, in general, $\chi^\pm$ becomes negative during the 
evolution, corresponding to the formation of a trapped surface.
 
\section{Local isometric embedding of surfaces}
\setcounter{equation}{0}

The mathematical theory of surfaces  was originally formulated \cite{RI1898} 
either by regarding surfaces as flexible but
inextendible layers, or by viewing them as having a rigid form in three-dimensional space.
What we are going to discuss hereafter is more directly related to the latter approach. 

Two surfaces $S,S'$ among which one can establish a 
correspondence such that their metrics turn out to be equal, have the same geometry. In that case also
finite arcs, angles and areas of figures upon $S$ are equal to their counterparts that correspond
to them upon $S'$. The two surfaces $S,S'$ are then said to be applicable one upon the other,
which means that, by means of a flexure, a surface (or at least a portion of a surface) can be unfolded
without any break nor duplication upon the other. For this folding to be feasible, it is clearly
necessary  to prove the existence of a continuous series of configurations of the flexible
surface $S$, that leads from $S$ to $S'$.

On denoting by 
\begin{equation}
g_{_{S}}=E {\rm d}u \otimes {\rm d}u+F ({\rm d}u \otimes {\rm d}v+{\rm d}v \otimes {\rm d}u)
+G {\rm d}v \otimes {\rm d}v
\label{(4.1)}
\end{equation}
the metric on the surface $S$, and by
\begin{equation}
g_{_{S'}}=E' {\rm d}u' \otimes {\rm d}u'
+F' ({\rm d}u' \otimes {\rm d}v'+{\rm d}v' \otimes {\rm d}u')
+G' {\rm d}v' \otimes {\rm d}v'
\label{(4.2)}
\end{equation}
the metric on the surface $S'$, the necessary and sufficient condition for the applicability of
$S$ and $S'$ is that $g_{_{S}}$ and $g_{_{S'}}$ can be transformed the one into the other
through an isometry, i.e.
\begin{equation}
g_{_{S}}=g_{_{S'}}.
\label{(4.3)}
\end{equation}
Once that two surfaces $S$ and $S'$ are given, the first applicability problem consists in
finding whether they can be applied the one upon the other and, in the affirmative case, in
obtaining the formulae ensuring the applicability.

The second (more important) problem of applicability theory consists in finding all surfaces 
applicable upon a given surface, i.e., in {\it finding all surfaces with a given metric}.
This problem was reduced to finding all solutions of a certain partial differential equation
(appendix C) by a highly original application of the theory of differential 
parameters \cite{DA3,BI1,BI2}, partial differential equations of the Monge-Amp\`ere form 
\cite{WE1896}, the method of characteristics \cite{WE1896}, the properties
of movable trihedra\footnote{Modern nomenclature refers to them as moving frames.} 
\cite{WE1896,BI2} and the absolute differential calculus \cite{RI1898}.

With modern language, we say that we try to understand whether a metric is 
smoothly realizable \cite{JA1971}. 
One has also to distinguish between local and global isometric embedding.
The theory of applicability was developed by the pioneering works of Darboux, Weingarten, Bianchi
and Ricci-Curbastro, for surfaces embedded in flat three-dimensional Euclidean space with metric
$$
{\rm d}x \otimes {\rm d}x+{\rm d}y \otimes {\rm d}y+{\rm d}z \otimes {\rm d}z,
$$
and even volumes $3$ and $5$ of a modern treatise such as the one by Spivak 
\cite{SP1979c, SP1979e}, and the book by Han and Hong \cite{HH2006}, 
focus on this sort of embedding. 
Within that framework, the Cartesian coordinates $(x,y,z)$ of any point on the surface
are taken to depend on two variables $(u,v)$. With any particular value $v_{1}$ of $v$ there corresponds
a special curve with parametric equations
$$
x=x(u,v_{1}), \qquad y=y(u,v_{1}), \qquad z=z(u,v_{1}),
$$
and, by varying $v$ in a continuous way, this curve undergoes a continuous motion in
$({\mathbb R}^{3},\delta_{E})$ and describes a surface, defined analytically by the equations
\begin{equation}
x=x(u,v), \qquad y=y(u,v), \qquad z=z(u,v).
\label{(4.4)}
\end{equation}
Among the $3$ equations (4.4), elimination of $u,v$ variables leads to a relation
(cf. Eqs. (4.4)) 
\begin{equation}
f(x,y,z)=0,
\label{(4.5)}
\end{equation}
which is the ordinary equation of the surface.

The second applicability problem was stated by Darboux as follows \cite{DA3}: if $E,F,G$ are 
given functions of the $(u,v)$ variables, find all functions $x(u,v),y(u,v),z(u,v)$  which satisfy
the equation
\begin{equation}
\fl\quad
{\rm d}x \otimes {\rm d}x+{\rm d}y \otimes {\rm d}y+{\rm d}z \otimes {\rm d}z
= E {\rm d}u \otimes {\rm d}u+F({\rm d}u \otimes {\rm d}v+{\rm d}v \otimes {\rm d}u)
+ G {\rm d}v \otimes {\rm d}v,
\label{(4.6)}
\end{equation}
where ${\rm d}u$ and ${\rm d}v$ can be taken arbitrarily. Darboux pointed out that Eq. (4.6)
can be re-expressed in the form
\begin{eqnarray}
\fl\quad
{\rm d}x \otimes {\rm d}x+{\rm d}y \otimes {\rm d}y
&=& E {\rm d}u \otimes {\rm d}u+F({\rm d}u \otimes {\rm d}v+{\rm d}v \otimes {\rm d}u)
+G {\rm d}v \otimes {\rm d}v 
\nonumber \\
\fl\quad
&-& \left({\partial z \over \partial u}{\rm d}u+{\partial z \over \partial v}{\rm d}v \right)
\otimes \left({\partial z \over \partial u}{\rm d}u
+{\partial z \over \partial v}{\rm d}v \right).
\label{(4.7)}
\end{eqnarray}
Of course, nothing really changes if, within this scheme, we bring on the right-hand side
${\rm d}y \otimes {\rm d}y$ or ${\rm d}x \otimes {\rm d}x$ instead of ${\rm d}z \otimes {\rm d}z$.
In this equation, the left-hand side is the Euclidean metric in the $(x,y)$ plane, which is flat. 
Thus, following Darboux one has to impose that the curvature pertaining to the metric on the
right-hand side of Eq. \eqref{(4.7)} must vanish as well. 
After some non-trivial rearrangements of the several
terms involved in the calculation, he arrived at a second-order non-linear partial differential equation 
for the function $z$, of the Monge-Amp\`ere type (see now appendix C for a more rigorous formulation
of the Darboux method).

The previous outline prepares the ground for our analysis. In our case, the surface is embedded in
a three-dimensional curved Riemannian manifold, and we can no longer consider 
Cartesian coordinates of points on the surface. 
However, we can still write down \lq\lq space coordinates" $Y^{j}$ $(j=1,2,3)$ (hereafter 
$Y^1=\xi$, $Y^2=\eta$, $Y^3=\zeta$ to follow standard notation) and \lq\lq surface coordinates" 
$X^a$ $(a=1,2)$ (hereafter $X^1=u$, $X^2=v$ to follow standard notation), i.e.
\begin{equation}
Y^1=\xi(u,v),\quad Y^2=\eta(u,v),\quad Y^3=\zeta(u,v)\,,
\label{(4.8)}
\end{equation}
with
\begin{equation}
X^1=u \qquad X^2=v\,.
\label{(4.9)}
\end{equation}
The metric on the left-hand side of Eq. \eqref{(4.6)} is then replaced by the Riemannian metric
(see appendix B)
\begin{equation}
h= \sum_{i,j=1}^{3}h_{ij}dY^{i} \otimes dY^{j},
\label{(4.10)}
\end{equation}
where
\begin{equation}\fl\qquad
h_{ij}|_{\Sigma_2}=h_{ij}(Y^1,Y^2,Y^3)=h_{ij}(\xi(u,v),\eta(u,v),\zeta(u,v))\equiv \tilde h_{ij}(u,v). 
\label{(4.11)}
\end{equation}
We therefore state the applicability problem when $(\Sigma_{2},H)$ is embedded into a Riemannian
$3$-manifold as follows: if $E,F,G$ are {\it given} functions of some $(u,v)$ variables, and hence
\begin{equation}
H=\sum_{a,b=1}^{2} H_{ab}dX^{a}\otimes dX^{b}
\label{(4.12)}
\end{equation}
is a {\it given} two-dimensional Riemannian metric, find all
functions $\xi,\eta,\zeta$ of $(u,v)$ such that
\begin{equation}
h|_{\Sigma_2}=E {\rm d}u \otimes {\rm d}u+F({\rm d}u \otimes {\rm d}v+{\rm d}v \otimes {\rm d}u)
+G {\rm d}v \otimes {\rm d}v,
\label{(4.13)}
\end{equation}
i.e.,
\begin{eqnarray}
\fl\qquad
\sum_{i,j=1}^{3} h_{ij}dY^{i} \otimes dY^{j}|_{\Sigma_2}
&=&\sum_{i,j=1}^{3} \sum_{a,b=1}^{2}
\tilde h_{ij}\frac{\partial Y^i}{\partial X^a}
\frac{\partial Y^j}{\partial X^b}dX^{a}\otimes dX^{b}\equiv H ,
\label{(4.14)}
\end{eqnarray}
with $H$ defined in Eq. \eqref{(4.12)}.
The non-trivial problem consists in identifying the embedding functions $Y^{j}(X^{a})$. 
Since any direct approach is rather involved, we procced instead by generalizing as follows 
the method due to Darboux.

This means that, following Darboux \cite{DA3}, 
we are led to re-express Eq. \eqref{(4.13)} in the form (with our notation,
${ }^{(2)}f$ is the metric of a flat $2$-dimensional space, and we write in
square brackets the association among local coordinates)
\begin{eqnarray}
\fl\quad
{ }^{(2)}f[\xi \leftrightarrow \xi, \eta \leftrightarrow \eta] &=&   
{\rm d}\xi \otimes {\rm d}\xi + {\rm d}\eta \otimes {\rm d}\eta 
\nonumber \\
&=& {1 \over h_{11}}\biggr[E {\rm d}u \otimes {\rm d}u 
+F ({\rm d}u \otimes {\rm d}v+{\rm d}v \otimes {\rm d}u)
+G {\rm d}v \otimes {\rm d}v 
\nonumber \\
\fl\quad
&-& (h_{22}-h_{11}){\rm d}\eta \otimes {\rm d}\eta -h_{33}{\rm d}\zeta \otimes {\rm d}\zeta
\nonumber \\
\fl\quad
&-& h_{12}({\rm d}\xi \otimes {\rm d}\eta + {\rm d}\eta \otimes {\rm d}\xi)
-h_{23}({\rm d}\eta \otimes {\rm d}\zeta + {\rm d}\zeta \otimes {\rm d}\eta)
\nonumber \\
\fl\quad
&-& h_{31}({\rm d}\zeta \otimes {\rm d}\xi + {\rm d}\xi \otimes {\rm d}\zeta)\biggr].
\label{(4.15)}
\end{eqnarray}
At this stage, it is still true that we arrive at an equation where the left-hand side
is the metric on a (fictitious) $(\xi,\eta)$ plane with vanishing curvature, and hence we
can again require that the curvature of the metric on the right-hand side of Eq. \eqref{(4.15)}
must vanish, where
\begin{equation}
{\rm d}\Gamma=\sum_{a=1}^{2} \frac{\partial \Gamma}{\partial X^a}dX^{a}
={\partial \Gamma \over \partial u}{\rm d}u+{\partial \Gamma \over \partial v}{\rm d}v, \qquad
\forall \Gamma =\xi,\eta,\zeta.
\label{(4.16)}
\end{equation}
Of course, the task of exploiting Eq. \eqref{(4.15)} to obtain the equations for applicability is
harder than the one relying upon Eq. \eqref{(4.7)}, because now all differentials
${\rm d}\xi,{\rm d}\eta$ and ${\rm d}\zeta$ occur on the right-hand side.
Nevertheless, one can make further progress by writing other two equations obtainable from
the above by permutation, i.e.
\begin{eqnarray}
\fl\quad
{ }^{(2)}f[\eta \leftrightarrow \eta, \zeta \leftrightarrow \zeta]&=&  
{\rm d}\eta \otimes {\rm d}\eta + {\rm d}\zeta \otimes {\rm d}\zeta 
\nonumber \\
\fl\quad
&=& {1 \over h_{22}}\biggr[E {\rm d}u \otimes {\rm d}u 
+F ({\rm d}u \otimes {\rm d}v+{\rm d}v \otimes {\rm d}u)
+G {\rm d}v \otimes {\rm d}v 
\nonumber \\
\fl\quad
&-& (h_{33}-h_{22}){\rm d}\zeta \otimes {\rm d}\zeta -h_{11}{\rm d}\xi \otimes {\rm d}\xi
\nonumber \\
\fl\quad
&-& h_{12}({\rm d}\xi \otimes {\rm d}\eta + {\rm d}\eta \otimes {\rm d}\xi)
-h_{23}({\rm d}\eta \otimes {\rm d}\zeta + {\rm d}\zeta \otimes {\rm d}\eta)
\nonumber \\
\fl\quad
&-& h_{31}({\rm d}\zeta \otimes {\rm d}\xi + {\rm d}\xi \otimes {\rm d}\zeta)\biggr],
\label{(4.17)}
\end{eqnarray}
\begin{eqnarray}
\fl\quad
{ }^{(2)}f[\zeta \leftrightarrow \zeta, \xi \leftrightarrow \xi] &=& 
{\rm d}\zeta \otimes {\rm d}\zeta + {\rm d}\xi \otimes {\rm d}\xi 
\nonumber \\
\fl\quad
&=& {1 \over h_{33}}\biggr[E {\rm d}u \otimes {\rm d}u 
+F ({\rm d}u \otimes {\rm d}v+{\rm d}v \otimes {\rm d}u)
+G {\rm d}v \otimes {\rm d}v 
\nonumber \\
\fl\quad
&-& (h_{11}-h_{33}){\rm d}\xi \otimes {\rm d}\xi -h_{22}{\rm d}\eta \otimes {\rm d}\eta
\nonumber \\
\fl\quad
&-& h_{12}({\rm d}\xi \otimes {\rm d}\eta + {\rm d}\eta \otimes {\rm d}\xi)
-h_{23}({\rm d}\eta \otimes {\rm d}\zeta + {\rm d}\zeta \otimes {\rm d}\eta)
\nonumber \\
\fl\quad
&-& h_{31}({\rm d}\zeta \otimes {\rm d}\xi + {\rm d}\xi \otimes {\rm d}\zeta)\biggr],
\label{(4.18)}
\end{eqnarray}
which should be supplemented by a triplet of equations, whose left-hand sides are 
the $2$-metrics
$$
\Bigr(d\xi \otimes d \eta+ d\eta \otimes d \xi \Bigr), \qquad
\Bigr(d \eta \otimes d \zeta + d \zeta \otimes d \eta \Bigr), \qquad
\Bigr(d \zeta \otimes d\xi + d \xi \otimes d \zeta \Bigr),
$$
respectively, which also pertain to a plane with vanishing curvature. These additional
equations read as
\begin{eqnarray}
\fl\quad
{ }^{(2)}f[\xi \leftrightarrow \eta, \eta \leftrightarrow \xi] &=& 
\Bigr(d\xi \otimes d \eta + d\eta \otimes d\xi \Bigr) 
\nonumber \\
\fl\quad
&=&{1 \over h_{12}}\biggr[E du \otimes du+F (du \otimes dv + dv \otimes du)
+G dv \otimes dv \nonumber \\
\fl\quad
&-& h_{11}d\xi \otimes d\xi -h_{22} d\eta \otimes d\eta
-h_{33}d\zeta \otimes d\zeta 
\nonumber \\
\fl\quad
&-& h_{23}(d\eta \otimes d\zeta+d\zeta \otimes d\eta)
-h_{31}(d\zeta \otimes d\xi + d\xi \otimes d\zeta)\biggr],
\label{(4.19)}
\end{eqnarray}
\begin{eqnarray}
\fl\quad
{ }^{(2)}f[\eta \leftrightarrow \zeta, \zeta \leftrightarrow \eta] &=& 
\Bigr(d\eta \otimes d \zeta + d\zeta \otimes d\eta \Bigr) 
\nonumber \\
\fl\quad
&=&{1 \over h_{23}}\biggr[E du \otimes du+F (du \otimes dv + dv \otimes du)
+G dv \otimes dv \nonumber \\
\fl\quad 
&-& h_{11}d\xi \otimes d\xi -h_{22} d\eta \otimes d\eta
- h_{33}d\zeta \otimes d\zeta 
\nonumber \\
\fl\quad
&-& h_{12}(d\xi \otimes d\eta+d\eta \otimes d\xi)
-h_{31}(d\zeta \otimes d\xi + d\xi \otimes d\zeta)\biggr],
\label{(4.20)}
\end{eqnarray}
\begin{eqnarray}
\fl\quad
{ }^{(2)}f[\zeta \leftrightarrow \xi, \xi \leftrightarrow \zeta] &=& 
\Bigr(d\zeta \otimes d \xi + d\xi \otimes d\zeta \Bigr) 
\nonumber \\
\fl\quad
&=&{1 \over h_{31}}\biggr[E du \otimes du+F (du \otimes dv + dv \otimes du)
+G dv \otimes dv 
\nonumber \\
\fl\quad
&-& h_{11}d\xi \otimes d\xi -h_{22} d\eta \otimes d\eta
- h_{33}d\zeta \otimes d\zeta 
\nonumber \\
\fl\quad
&-& h_{12}(d\xi \otimes d\eta+d\eta \otimes d\xi)
-h_{23}(d\eta \otimes d\zeta + d\zeta \otimes d\eta)\biggr]\,,
\label{(4.21)}
\end{eqnarray}
where on the right-hand-side one should express $\xi$, $\eta$, $\zeta$ as functions of $u$ and $v$.

Thus, the isometric embedding of a surface into a curved three-dimensional Riemannian manifold 
makes it necessary to study, in general, 
a coupled set of $6$ partial differential equations for
the unknown functions $\xi,\eta,\zeta$, if one wants to solve the second applicability problem.

\section{The case of FLRW cosmologies}
\setcounter{equation}{0}

The formulae of the previous section are interesting because they provide a method for
extending the Darboux method when the Euclidean $3$-space is replaced by a generic 
three-dimensional Riemannian manifold. However, the resulting vanishing curvature 
equations are virtually intractable, and one has to resort to a more specific class
of Riemannian $3$-manifolds. Interestingly, the Friedmann-(Lemaitre)-Robertson-Walker 
cosmologies of general relativity provide a valuable example, because for them the
space-time metric with scale factor $a(t)$ can be written in the form \cite{HE1973}
\begin{equation}
\fl\qquad\qquad
g=-dt \otimes dt+a^{2}(t)\Bigr[d\chi \otimes d\chi
+f^{2}(\chi) \Bigr(d\theta \otimes d\theta+\sin^{2}\theta d\phi \otimes d\phi \Bigr)
\Bigr],
\label{(5.1)}
\end{equation}
where, denoting by $\kappa$ the suitably normalized curvature, one has
\begin{equation}
\label{5.234}
f(\chi)=\left\{ 
\begin{array}{ll}
\sin \chi &\qquad   \kappa=\phantom{-}1, \;\chi \in [0,2 \pi]
\cr 
\chi &\qquad   \kappa=\phantom{-}0, 
\; \chi \in {\mathbb R}^{+} \cup \{ 0 \}
\cr 
\sinh \chi &\qquad   \kappa=-1.  
\;  \chi \in {\mathbb R}^{+} \cup \{ 0 \}
\end{array}
\right.
\end{equation}
The condition \eqref{(4.14)} for local isometric embedding reads therefore as
\begin{eqnarray}
\; & \; &
a^{2} \Bigr[d\chi \otimes d\chi+f^{2}(\chi)(d\theta \otimes d\theta
+\sin^{2} \theta d\phi \otimes d\phi)\Bigr]
\nonumber \\
&=& E du \otimes du+F (du \otimes dv+dv \otimes du )
+G dv \otimes dv .
\label{(5.3)}
\end{eqnarray}
At this stage, inspired by the work in Ref. \cite{DING1999}, we define a new variable $w$
by requiring that
\begin{equation}
e^{w}=a(t)f(\chi) \quad\Longrightarrow\quad 
d\chi={e^{w}\over a}{dw \over f'(\chi(w))}
={f(\chi(w))\over f'(\chi(w))}dw,
\label{(5.4)}
\end{equation}
and hence, from Eqs. \eqref{(5.3)} and \eqref{(5.4)}, we find the relation
\begin{eqnarray}
\fl\quad
\Bigr[E du \otimes du+F (du \otimes dv + dv \otimes du)
+G dv \otimes dv \Bigr]e^{-2w}-{{dw \otimes dw} \over [f'(\chi(w))]^{2}}
\nonumber \\
\fl\quad
= d \theta \otimes d \theta +\sin^{2}\theta d\phi \otimes d\phi.
\label{(5.5)}
\end{eqnarray}
The metric on the right-hand side of Eq. \eqref{(5.5)} is the metric on the unit $2$-sphere,
for which the Gaussian curvature is $+1$. From the left-hand side of Eq. (5.5), we
obtain $3$ different realizations of this positive curvature condition, depending on
whether we take $f(\chi)$ from the first, second or third of 
Eqs. \eqref{5.234}. 

\subsection{The spatially flat case}

The case $f(\chi)=\chi$, corresponding to a spatially flat Friedmann-Robertson-Walker
universe, is the less difficult. For it $f'(\chi)=1$, and we obtain from Eq. (5.5) the
partial differential equation (cf. Eq. (2.9) in Ref. \cite{DING1999}, which is instead
a negative Gauss curvature condition)
\begin{equation}
1={1 \over (1-{\tilde {\bigtriangleup}}_{1}w )}
\left[{\tilde K} - {{\tilde {\bigtriangleup}}_{22}w \over
(1-{\tilde {\bigtriangleup}}_{1}w)} \right],
\label{(5.6)}
\end{equation}
where the differential parameters ${\tilde {\bigtriangleup}}_{1}$ and
${\tilde {\bigtriangleup}}_{22}$ are built from the conformally rescaled metric 
${\tilde g}$ on the left-hand side of (5.5), with covariant form
\begin{equation}
{\tilde g}_{ij}= 
\left(
\begin{array}{cc}
E & F \cr 
F & G \cr
\end{array}
\right) e^{-2w}.
\label{(5.7)}
\end{equation}
The formula (5.6) is therefore interpreted as follows.

${\tilde {\bigtriangleup}}_{1}$ is the first differential parameter \cite{BI1} of 
$w=w(u^{1},u^{2})=w(u,v)$ with respect to the metric ${\tilde g}$, i.e.
\begin{equation}
{\tilde {\bigtriangleup}}_{1}=\sum_{i,j=1}^{2}
{\tilde g}^{ij}w_{,i} \; w_{,j}
=\langle {\rm grad}w,{\rm grad}w \rangle_{{\tilde g}},
\label{(5.8)}
\end{equation}
while ${\tilde K}$ is the Gauss curvature pertaining to ${\tilde g}$ and
${\tilde {\bigtriangleup}}_{22}$ is the second-order differential parameter \cite{BI1}
\begin{equation}
{\tilde {\bigtriangleup}}_{22}w ={({\tilde {\nabla}}_{uu}w) ({\tilde {\nabla}}_{vv}w)
-({\tilde {\nabla}}_{uv}w)^{2} \over {\rm det}({\tilde g})},
\label{(5.9)}
\end{equation}
with the understanding that covariant derivaties with connection coefficients from
the metric ${\tilde g}$ read as ($i,j=1,2; \; f_{i} \equiv w_{,i}$)
\begin{equation}
{\tilde {\nabla}}_{j}f_{i}=f_{i,j}
-{\tilde \Gamma}_{\; ij}^{1} \; f_{1}
-{\tilde \Gamma}_{\; ij}^{2}f_{2}.
\label{(5.10)}
\end{equation}

The equation (5.6) to be solved in order to prove local isometric embedding 
reads eventually as
\begin{equation}
{\tilde {\bigtriangleup}}_{22}w={\tilde K} \Bigr(1-{\tilde {\bigtriangleup}}_{1}w \Bigr)
-\Bigr(1-{\tilde {\bigtriangleup}}_{1}w \Bigr)^{2}.
\label{(5.11)}
\end{equation}
Unlike the Darboux equation (C.9) or (C.20) for the local isometric embedding of a surface
into Euclidean $3$-space, Eq. (5.11) contains on the right-hand side also the 
additional term $\Bigr(1-{\tilde {\bigtriangleup}}_{1}w \Bigr)^{2}$. 
Now we must insert into Eq. (5.11) the components $E,F,G$ of the first fundamental 
form of a trapped surface in order to solve our problem. This task is accomplished in the 
following section.

\section{Proof of realizability of trapped surfaces in FLRW cosmologies}

Once the metric of a FLRW space-time is written in the form \eqref{(5.1)}, the null geodesic 
vector fields of Eq. (3.19) read as
\begin{equation}
l^{\pm}={1 \over a(t)}{\partial \over \partial t} \pm 
{1 \over a^{2}(t)}{\partial \over \partial r},
\label{(6.1)}
\end{equation}
and hence the space-time tensor field defined in Eq. (2.4) reduces to
(see also Ref. \cite{Casadio})
\begin{equation}
h=\sum_{\mu,\nu=1}^{4} h_{\mu \nu}dx^{\mu} \otimes dx^{\nu}
=\sum_{i,j=1}^{2}H_{ij}d\zeta^{i} \otimes d\zeta^{j},
\label{(6.2)}
\end{equation}
which is therefore the metric $H$ of a $2$-surface, our trapped surface. In
the spatially flat case, where $f(\chi)=\chi$, we can write
\begin{equation}
H=a^{2}(t)\chi^{2}(d\theta \otimes d\theta 
+ \sin^{2} \theta d \phi \otimes d\phi). 
\label{(6.3)}
\end{equation}
Moreover, the function $w=w(u,v)$ of Sec. $5$ is here a function of the 
variables $\theta$ and $\phi$, and the $2$-metric $\gamma$ equal to the whole left-hand 
side of Eq. (5.5) has $F=0,E=a^{2}\chi^{2},G=a^{2}(t)\chi^{2}\sin^{2}\theta$ from
\eqref{(6.3)}, while $w=w(\theta,\phi)$. 
Hence we find
\begin{equation}
\gamma=\gamma_{11}d\theta \otimes d\theta
+\gamma_{12}(d\theta \otimes d\phi + d\phi \otimes d\theta)
+\gamma_{22}d\phi \otimes d\phi,
\label{(6.4)}
\end{equation}
where the matrix of components reads as
\begin{equation}
\gamma_{ij}=\left(\begin{array}{cc}
1-(w_{,\theta})^{2} & -w_{,\theta} \; w_{,\phi} \\
-w_{,\theta} \; w_{,\phi} & \sin^{2}\theta-(w_{,\phi})^{2}
\end{array}\right),
\label{(6.5)}
\end{equation}
having set $w_{,\theta} \equiv {\partial w \over \partial \theta}, \;
w_{,\phi} \equiv {\partial w \over \partial \phi}$. Now,  following Sec. $5$,   
we require that the Gauss curvature pertaining to the metric \eqref{(6.5)} should be
equal to $1$. The form taken by Eq. \eqref{(5.8)} 
in the $(\theta,\phi)$ 
coordinates is cumbersome, but if we assume that
\begin{equation}
w(\theta,\phi)=W(\theta),
\label{(6.6)}
\end{equation}
i.e. a function depending only on the $\theta$ variable, we obtain  the
non-linear ordinary differential equation
\begin{equation}
W''(\theta)= \tan \theta W'(\theta)\Bigr[1-(W'(\theta))^{2}\Bigr],
\label{(6.7)}
\end{equation}
which is solved by
\begin{equation}
W(\theta)=C_2\pm \frac{1}{\sqrt{2(1+\cos\theta)}}
F_{J}(\cos \theta,C_{1}) ,
\label{(6.8)}
\end{equation}
having denoted by $C_{1},C_{2}$ two integration constants and by 
$F_{J}(z,k)$ Jacobi's form (hence the subscript $J$) 
of the elliptic integral of first kind
\begin{equation}
F_{J}(z,k)=\int_{0}^{z} {dt \over \sqrt{1-t^{2}} \;
\sqrt{1-k^{2}t^{2}}}.
\label{(6.9)}
\end{equation}
We display in figure $1$ the particular plot of $W(\theta)$ obtained upon choosing
the values $C_{1}=1,C_{2}=0$ for the constants.

\begin{figure}
\begin{center}
\includegraphics[scale=0.35]{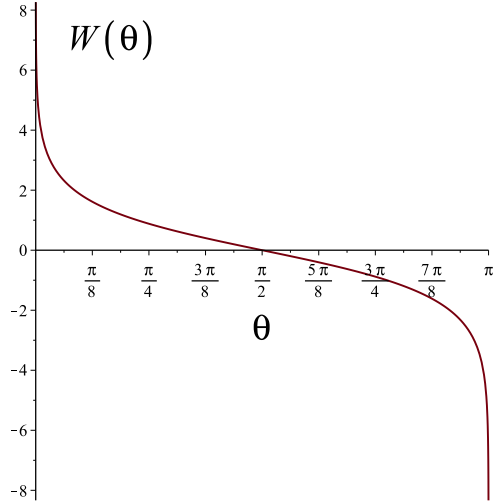}
\end{center}
\caption{The behaviour  of $W(\theta)$, solution 
\eqref{(6.8)} of Eq. \eqref{(6.7)}, is shown  for 
the values $C_{1}=1,C_{2}=0$ of the integration constants.}
\end{figure}

For a closed or open FLRW universe, respectively, the method remains the same but 
the curvature calculation for the metric equal to the left-hand side of Eq. (5.5) is 
a bit more cumbersome. In the closed or open FLRW universe one has again the null
geodesic vector fields expressed in the form (6.1), while the $2$-metric $H$ in the
closed FLRW model is
\begin{equation}
H=a^{2}(t)\sin^{2}\chi (d \theta \otimes d \theta 
+ \sin^{2} \theta d \phi \otimes d \phi).
\label{(6.10)}
\end{equation}
The resulting metric given by the sum of all terms on the left-hand side of 
Eq. (5.5) is now described by the $2 \times 2$ matrix
\begin{equation}
\gamma_{ij}=\left(\begin{array}{cc}
1-{(w_{,\theta})^{2} \over (1-a^{-2}(t)e^{2w})} & 
-{w_{,\theta} \; w_{,\phi} \over (1-a^{-2}(t)e^{2w})} \\
-{w_{,\theta} \; w_{,\phi} \over (1-a^{-2}(t)e^{2w})} &
\sin^{2}\theta -{(w_{,\phi})^{2} \over (1-a^{-2}(t)e^{2w})}
\end{array}\right).
\label{(6.11)}
\end{equation}
The resulting curvature calculation consists of several lines, but the search
for a function $w(\theta,\phi)=W(\theta)$ as before leads to a helpful breakthrough,
because the function $W(\theta)$ is found to obey the non-linear equation
\begin{equation}
W''(\theta)= \tan \theta W'(\theta)
\left[1+a^{2}{(W'(\theta))^{2}\over (e^{2W}-a^{2})}\right]
+{(W'(\theta))^{2} e^{2W}\over (e^{2W}-a^{2})}.
\label{(6.12)}
\end{equation}
The general solution involves again the incomplete elliptic integral (6.9) and two
integration constants $C_{1},C_{2}$. We plot in figure $2$ the   particular 
function $W(\theta)$ obtained upon choosing $C_{1}=1$ and $C_{2}=0$, which has
explicit form (cf. Ref. \cite{Malec})
\begin{equation}
W(\theta)=-\ln\left(\cosh\left(\frac14 
\ln\left(\frac{(\cos(\theta)+1)^2}{(\cos(\theta)-1)^2}\right)\right)\right).
\label{(6.13)}
\end{equation}

\begin{figure}
\begin{center}
\includegraphics[scale=0.35]{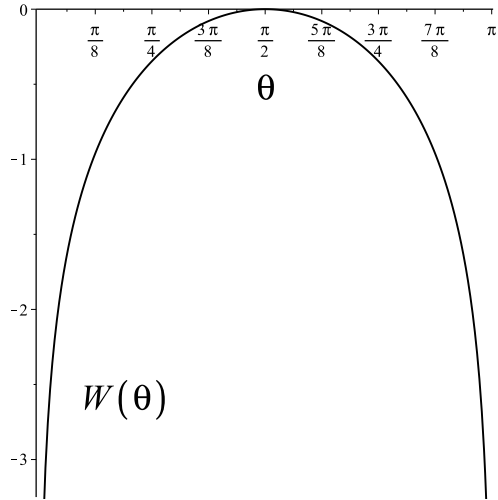}
\end{center}
\caption{The behaviour  of $W(\theta)$, given in  Eq. \eqref{(6.13)}, is shown  for 
the values $C_{1}=1,C_{2}=0$ of the integration constants.}
\end{figure}

Last, in the hyperbolic FLRW case, one deals with the
trapped-surface-$2$-metric
\begin{equation}
H=a^{2}(t)\sinh^{2}\chi (d\theta \otimes d\theta
+\sin^{2}\theta d\phi \otimes d\phi).
\label{(6.14)}
\end{equation}
The resulting metric equal to the whole left-hand side of Eq. (5.5) is described by
the $2 \times 2$ matrix of components
\begin{equation}
\gamma_{ij}=\left(\begin{array}{cc}
1-{(w_{,\theta})^{2} \over (1+a^{-2}(t)e^{2w})} & 
-{w_{,\theta} \; w_{,\phi} \over (1+a^{-2}(t)e^{2w})} \\
-{w_{,\theta} \; w_{,\phi} \over (1+a^{-2}(t)e^{2w})} &
\sin^{2} \theta -{(w_{,\phi})^{2} \over (1+a^{-2}(t)e^{2w})}
\end{array}\right).
\label{(6.15)}
\end{equation}
Once more, the resulting curvature calculation in the $(\theta,\phi)$ 
coordinates is feasible, and if the assumption (6.6) is made we find for
$W(\theta)$ the non-linear ordinary differential equation
\begin{equation}
W''(\theta)= \tan \theta W'(\theta)
\left[1-a^{2}{(W'(\theta))^{2}\over (e^{2W}+a^{2})}\right]
+{(W'(\theta))^{2} e^{2W} \over (e^{2W}+a^{2})}\,.
\label{(6.16)}
\end{equation}
In figure $3$ we plot the particular function $W(\theta)$ obtained upon choosing
the integration constants $C_{1}=C_{2}=1$, and having the explicit form
\begin{equation}
W(\theta)={1 \over 2}\log {4 \over 
(e^{(2-2 {\rm arctanh}(\cos \theta))}-1)^{2}}
-{\rm arctanh}(\cos \theta)+1.
\label{(6.17)}
\end{equation}

\begin{figure}
\begin{center}
\includegraphics[scale=0.35]{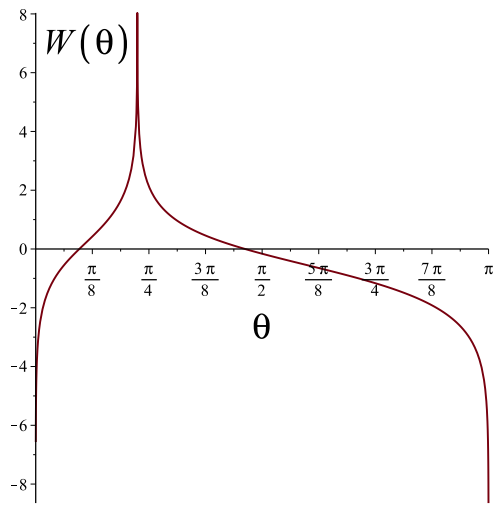}
\end{center}
\caption{The behaviour  of $W(\theta)$, given in  Eq. \eqref{(6.17)}, is shown  for 
the values $C_{1}=1,C_{2}=1$ of the integration constants. Note that a different choice 
of the integration constants allows for a shift of the point where the divergence of 
$W(\theta)$ occurs. However, the fact that the $\theta$-interval may not be maximal 
imposes \lq\lq pathologies" to the solution which can be eventually ruled out as being 
\lq\lq spurious" or \lq\lq non-physical."  
}
\end{figure}

The final consistency check for all our particular forms of $W(\theta)$ is that the form 
of $\chi$ resulting from the definition \eqref{(5.4)}, i.e. $f(\chi)={e^{W(\theta)}\over a}$, 
should be defined precisely on the intervals in \eqref{5.234}, or on proper subsets of
such intervals.

We conclude this section by mentioning the possibility of analyzing the dependence of the 
solutions $W(\theta)$ on the length-scale parameter $a$, $W=W_a(\theta)$ in both cases of open 
and closed universes (in the flat case the equation for $W(\theta)$ does not depend on $a$).
In fact one can rewrite Eqs. \eqref{(6.12)} and \eqref{(6.16)}, respectively, in the form
\begin{eqnarray}
\fl\quad
W''(\theta)- \tan \theta W'(\theta)-(W'(\theta))^2 &=&\frac{a^2 (W'(\theta))^2}{(e^{2W}-a^2)}[1+\tan \theta W'(\theta)]
\nonumber\\
\fl\quad
W''(\theta)- \tan \theta W'(\theta)-(W'(\theta))^2 &=&-\frac{a^2 (W'(\theta))^2}{(e^{2W}+a^2)}[1+\tan \theta W'(\theta)],
\end{eqnarray}
where the second equation is also obtained from the first one by the replacement $a\to i \; a$.
Looking at these equations perturbatively (in $a$) one can  solve exactly the condition 
on the left-hand-side (for $W_0(\theta)$)
\begin{equation}
W_0''(\theta)- \tan \theta W_0'(\theta)-(W_0'(\theta))^2 =0,
\end{equation}
that is
\begin{equation}
W_0(\theta) = -\ln\left[C_1\ln\left(\frac{\cos \theta }{1+\sin \theta }\right)-C_2\right],
\end{equation}
and use then this solution on the right-hand-side to find the 
$O(a^2)$-corrections according to standard procedures.
We will not develop further this approach, which is beyond the scope of the present study.

\section{Concluding remarks and open problems}

Our work in Sections $4$, $5$ and $6$, inspired by Refs. \cite{DA3} and \cite{DING1999}, 
is original, because even the
modern mathematical literature \cite{JA1971,SP1979c,SP1979e,HH2006,AT2011,LIU2016}, has kept focusing
on Riemannian $2$-manifolds embedded into $3$-dimensional Euclidean space,
with the exception of the outstanding work in Ref. \cite{GUAN2017}.
Our paper is part of a research program aimed at considering both the non-linear
equations of the mathematical theory of surfaces and the non-linear nature of the Einstein
field equations. 

Interestingly, the non-linear partial differential equation to be solved in order to achieve 
local isometric embedding is a curvature condition, i.e., it has a clear 
geometrical meaning, expressing the vanishing curvature
of the Euclidean plane or the positive Gauss curvature of the unit $2$-sphere. 

While our progress in Section $4$ is only qualitative, our results in Sections $5$ and
$6$ are more substantial because we succeed in obtaining, for the first time as far as we 
know, some explicit forms of the unknown function $W(\theta)$ 
whose existence is required for us to be able to realize trapped surfaces in
cosmological backgrounds.

On the mathematical side, the investigation of Eq. \eqref{(5.11)} in suitable functional spaces
and for suitable initial data requires further work. On the physical side, the applications
of our method to other space-times of interest in general relativity might lead to a
better understanding of the trapping mechanism in the classical theory of the 
gravitational field.

\appendix

\section{Proof of the relation (2.13)}
\label{app1}

In order to prove Eq. (2.13) one has to show that the difference
$$
\chi^{+}-(g^{-1})^{\alpha \beta}\nabla_{\alpha}l^+_{\beta}
$$
vanishes identically. Indeed one finds
\begin{eqnarray}
\fl\quad
\chi^{+}-
(g^{-1})^{\alpha \beta}\nabla_{\alpha}l^+_{\beta}
&=&{1 \over 2}
l^{-}{}^{\beta}l^+{}^{\alpha}\nabla_{\alpha}l^+_{\beta}
+{1 \over 2}
l^+{}^{\beta}l^-{}^{\alpha}\nabla_{\alpha}l^+_{\beta}\nonumber\\
\fl\quad
&=& {1 \over 2} l^{-}{}^{\beta} \nabla_{l^{+}} l^+_{\beta}
+{1 \over 2}l^+{}^{\beta}\nabla_{l^{-}} l^+_{\beta}
\nonumber \\
\fl\quad
&=& {1 \over 2}
l^+{}^{\beta}l^-{}^{\alpha}\nabla_{\alpha}l^+_{\beta},
\label{(A.1)}
\end{eqnarray}
where in the last term we have used the geodesic condition for $l^+$,
\begin{equation}
\nabla_{l^+}l^+=0.
\label{(A.2)}
\end{equation}
Now note that by covariant differentiation one finds 
\begin{equation}
0=l^+{}^{\beta}\nabla_{\alpha}l^+_{\beta}
+l^+_{\beta}\nabla_{\alpha}l^+{}^{\beta},
\end{equation}
so that contracting by $l^-{}^{\alpha}$ 
\begin{eqnarray}
0&=&
l^+{}^{\beta}l^-{}^{\alpha}\nabla_{\alpha}l^+_{\beta}+l^-{}^{\alpha}l^+_{\beta}
\nabla_{\alpha}l^+{}^{\beta} 
= 2 
l^+{}^{\beta}l^-{}^{\alpha}\nabla_{\alpha}l^+_{\beta},
\label{(A.3)}
\end{eqnarray}
Using this condition in Eq. (\ref{(A.1)}) completes the proof.

\section{Immersion of surfaces}

The scheme considered in section $4$ is a particular case of the following geometric framework,
described in chapter $1$ of Ref. \cite{SP1979c}. If $M$ is a $m$-dimensional manifold, and $N$ is
a $q$-dimensional manifold, let
\begin{equation}
i: M \rightarrow N
\label{(B.1)}
\end{equation}
be the immersion of $M$ into $N$, and let us denote by $\langle \; , \; \rangle$ the Riemannian
metric of $N$. The induced Riemannian metric on $M$ is then $i^{*} \langle \; , \; \rangle$.
If $y^{1},...,y^{m+1}$ is a coordinate system on $N$
(hereafter $q=m+1$), with Riemannian metric
\begin{equation}
\langle \; , \; \rangle = \sum_{\alpha,\beta=1}^{m+1}G_{\alpha \beta}
{\rm d}y^{\alpha} \otimes {\rm d}y^{\beta},
\label{(B.2)}
\end{equation}
and if $x^{1},...,x^{m}$ is a coordinate system on the hypersurface $M$, one can
also express the metric (B.2) in the form
\begin{equation}
\langle \; , \; \rangle=\sum_{i,j=1}^{m}g_{ij}{\rm d}x^{i} \otimes {\rm d}x^{j}
\; {\rm on} \; M,
\label{(B.3)}
\end{equation}
for certain functions $g_{ij}$. The relation between $G_{\alpha \beta}$ and $g_{ij}$
is therefore as follows:
\begin{equation}
\sum_{\alpha,\beta=1}^{m+1}G_{\alpha \beta}
{\partial y^{\alpha}\over \partial x^{i}}
{\partial y^{\beta} \over \partial x^{j}}=g_{ij} \; {\rm on} \; M.
\label{(B.4)}
\end{equation}
The restriction of $y^{\alpha}$ to $M$ is again denoted by $y^{\alpha}$. In our section $4$,
$m=2$, so that $M$ is a surface, and $x^{1}=u,x^{2}=v$, while 
$y^{1}=\xi(u,v),y^{2}=\eta(u,v),y^{3}=\zeta(u,v)$. 
 
\section{Local isometric embedding of a surface into ${\mathbb R}^{3}$}

In modern language, as we already said in section $4$, 
applicability theory studies the purely local problem of isometrically
embedding a surface in ${\mathbb R}^{3}$. This means that we are given functions $g_{ij}=E,F,G$
on a neighbourhood of ${\mathcal O} \in {\mathbb R}^{2}$, with ${\rm det}(g_{ij})>0$, and we want to find a
function $f: U \rightarrow {\mathbb R}^{3}$, $U$ being a smaller neighbourhood of
${\mathcal O} \in {\mathbb R}^{2}$, such that the first fundamental form 
$I_{f}=f^{*}\langle \; , \; \rangle$ has components $g_{ij}=E,F,G$. Suppose first, as we do
in section 4, that we are given an immersion $f$ such that $I_{f}$ has components
$E,F,G$. Thus, we can write Eq. (4.6) where $(u,v)$ is, with the Darboux notation adopted here,
the standard coordinate system on ${\mathbb R}^{2}$. By composing the function $f$ with a Euclidean motion,
one can assume the conditions \cite{SP1979e}
\begin{equation}
z_{u}(0,0)=z_{v}(0,0)=0,
\label{(C.1)}
\end{equation}
which imply that
\begin{equation}
{\rm det} \left(
\begin{array}{cc}
 x_{u} & x_{v} \cr 
y_{u} & y_{v}
\end{array} \right) \not =0 \;\qquad {\rm at} \;\quad (u=0,v=0).
\label{(C.2)}
\end{equation}
On considering the tensor
\begin{eqnarray}
\langle \; , \; \rangle' & \equiv & E \; {\rm d}u \otimes {\rm d}u
+F \Bigr[{\rm d}u \otimes {\rm d}v + {\rm d}v \otimes {\rm d}u \Bigr]
+G \; {\rm d}v \otimes {\rm d}v-{\rm d}z \otimes {\rm d}z 
\nonumber \\
&=& \Bigr(E-(z_{u})^{2} \Bigr) {\rm d}u \otimes {\rm d}u
+(F-z_{u}z_{v})\Bigr[{\rm d}u \otimes {\rm d}v+{\rm d}v \otimes {\rm d}u \Bigr]
\nonumber \\
&+& \Bigr(G-(z_{v})^{2}\Bigr) {\rm d}v \otimes {\rm d}v,
\label{(C.3)}
\end{eqnarray}
the validity of Eq. (4.6) makes it possible to write
\begin{equation}
\langle \; , \; \rangle'={\rm d}x \otimes {\rm d}x
+{\rm d}y \otimes {\rm d}y.
\label{(C.4)}
\end{equation}
This is positive-definite at $(u=0,v=0)$ by virtue of (C.1), and hence positive-definite
in a neighbourhood of $(u=0,v=0)$. Moreover, $(x,y)$ is a coordinate system for ${\mathbb R}^{2}$
in a neighbourhood of $(u=0,v=0)$ by virtue of (C.2). Thus, Eq. (C.4) implies that the metric
$\langle \; , \; \rangle'$ is flat, and therefore has vanishing curvature $K'=0$.

Now one can prove by a direct calculation that the metric with coefficients $E,F,G$ 
has curvature $K$ given by \cite{SP1979c}
\begin{eqnarray}
\label{curvature}
\fl\qquad
K(EG-F^{2})&=& {\rm det} \left(
\begin{array}{ccc}
-{1 \over 2}G_{uu}+F_{uv}-{1 \over 2}E_{vv} & {1 \over 2} E_{u} 
& F_{u}-{1 \over 2}E_{v} \cr
F_{v}-{1 \over 2}G_{u} & E & F \cr
{1 \over 2}G_{v} & F & G 
\end{array}\right)
\nonumber \\
\fl\qquad
&-& {\rm det} \left(\begin{array}{ccc}
0 & {1 \over 2}E_{v} & {1 \over 2} G_{u} \cr
{1 \over 2}E_{v} & E & F \cr
{1 \over 2}G_{u} & F & G 
\end{array}\right).
\label{(C.5)}
\end{eqnarray}
In order to obtain the condition $K'=0$ one has to set to $0$ the difference of determinants on the
right-hand side of (C.5) while bearing in mind that we deal with the metric (C3), so that we have
to perform the replacements
\begin{equation}
E \rightarrow \Bigr(E-(z_{u})^{2}\Bigr), \;
F \rightarrow F-z_{u}z_{v}, \;
G \rightarrow \Bigr(G-(z_{v})^{2}\Bigr).
\label{(C.6)}
\end{equation}
With the Monge notation
\begin{equation}
p \equiv z_{u}, \; q \equiv z_{v},
\label{(C.7)}
\end{equation}
\begin{equation}
r \equiv z_{uu}, \; s \equiv z_{uv}, \; t \equiv z_{vv},
\label{(C.8)}
\end{equation}
the resulting {\it Darboux equation} for the vanishing of the curvature reads as
\begin{eqnarray}
&-&  4(EG-F^{2})(rt-s^{2}) 
\nonumber \\
&+& 2pr \Bigr[2GF_{v}-GG_{u}-FG_{v}\Bigr]
+2qr \Bigr[E G_{v}+FG_{u}-2FF_{v} \Bigr] 
\nonumber \\
&+& 4ps \Bigr[F G_{u}-G E_{v} \Bigr]+4qs \Bigr[F E_{v}-E G_{u} \Bigr]
\nonumber \\
&+& 2pt \Bigr[G E_{u}+F E_{v}-2 F F_{u}\Bigr]
+2qt \Bigr[2E F_{u}-EE_{v}-F E_{u} \Bigr]
\nonumber \\
&+& (E-p^{2})\Bigr[E_{v}G_{v}-2F_{u}G_{v}+(G_{u})^{2}\Bigr]
\nonumber \\
&+& (F-pq)\Bigr[E_{u}G_{v}-E_{v}G_{u}-2E_{v}F_{v}-2G_{u}F_{u}+4F_{u}F_{v}\Bigr]
\nonumber \\
&+& (G-q^{2})\Bigr[G_{u}E_{u}-2F_{v}E_{u}+(E_{v})^{2}\Bigr]
\nonumber \\
&+& 2 \Bigr[EG-F^{2}-Gp^{2}-Eq^{2}+2Fpq \Bigr]
\Bigr[2F_{uv}-E_{vv}-G_{uu}\Bigr]=0.
\label{(C.9)}
\end{eqnarray}
The Darboux equation is a non-linear partial differential equation 
for $z=z(u,v)$, but it is linear in the Jacobian $(rt-s^{2})$ and in the second
derivatives $r,s,t$. It belongs therefore to the 
family of Monge-Amp\`ere equations, and can be written in the form
\begin{equation}
t(r+Ap+Bq)+C=0,
\label{(C.10)}
\end{equation}
where $A,B,C$ do not involve $t=z_{vv}$, and hence one can solve for $t$ according to
\begin{equation}
t=-{C \over (r+Ap+Bq)}=t(u,v,p,q,r,s).
\label{(C.11)}
\end{equation}
This means that, given initial conditions along the $u$-axis such that the denominator
$(r+Ap+Bq)$ does not vanish at $(u=0,v=0)$, then one can write the equation for $z_{vv}$
in the form (C.11) near $(u=0,v=0)$. Interestingly, the Cauchy problem for such an equation is 
equivalent to the Cauchy problem for a quasi-linear first-order system, which is always
solvable, by virtue of the Cauchy-Kowalewski theorem, if all functions in the equation and all
initial data are analytic. Thus, the desired isometric embedding $f$ exists locally if 
$E,F,G$ are analytic \cite{SP1979e}.

Moreover, if a surface $M \subset {\mathbb R}^{3}$ has curvature $K>0$ everywhere and a metric which is
analytic in {\it some} coordinate system, then $M$ is actually an analytic submanifold
of ${\mathbb R}^{3}$. In particular, if $M$ is any surface of class $C^{3}$ of constant positive 
curvature, then $M$ is analytic. However, when $E,F,G$ are not analytic,
{\it there is no known criterion on the initial data} that guarantees the existence of a
solution of the Darboux equation (C.9). On the other hand, the lack of a sufficient condition
does not prevent us from proving the existence of some solutions, and this has been achieved
by Jacobowitz \cite{JA1971}.

When the curvature $K$ of $M$ is negative, one can prove that for any initial conditions with
$$
F_{t} \equiv r +Ap+Bq \not =0
$$
there always exists a solution $z$ of the Darboux Eq. (C.9) in a neighbourhood of
$(u=0,v=0)$, and the resulting solutions are less differentiable than the functions
$E,F,G$ \cite{SP1979e}.

Let us complete this section by showing that Eq. \eqref{curvature} allows for 
a simple geometrical  interpretation.
For this purpose let us consider the two-dimensional metrics $g_0$  and $g$ given by
\begin{equation}
ds_0^2=g^0_{\alpha\beta}dx^\alpha dx^\beta= E  du^2 +2 F dudv + G dv^2 ,
\label{(C.12)}
\end{equation}
and
\begin{eqnarray}
ds^2 &=& g_{\alpha\beta}dx^\alpha dx^\beta\nonumber\\
&=&(E-(z_{u})^{2}) du^2 +2(F-z_{u}z_{v})dudv +(G-(z_{v})^{2})dv^2\nonumber\\
&=& (g^0_{\alpha\beta}-z_\alpha z_\beta) dx^\alpha dx^\beta .
\label{(C.13)}
\end{eqnarray}
Let $R(g^0)$ and  $R(g)$ be the Ricci scalars associated with the two metrics, respectively, 
and define the following forms
\begin{equation}
dz=(\partial_\alpha z) dx^\alpha =z_u du +z_v dv , 
\label{(C.14)}
\end{equation}
and  
\begin{equation}
(\nabla^2  z)_{\alpha\beta}= \nabla_\alpha \partial_\beta z .
\label{(C.15)}
\end{equation}
Consider the quantities
\begin{equation}
\fl\quad
\nabla_1 z =g^{(0)}{}^{\alpha\beta} \partial_\alpha z  \partial_\beta z\,,\qquad  \nabla_2 z
=(\nabla^2  z)_{uu}  (\nabla^2  z)_{vv} -(\nabla^2  z)_{uv}^{2} .
\label{(C.16)}
\end{equation}
The following identity holds\footnote{Our convention for the definition of Ricci tensor is 
$R_{\alpha\beta}=g^{\mu\nu}R_{\mu\alpha\nu\beta}$.}
\begin{equation}
\frac{2 \nabla_2 z}{(E G-F^2)}- (-1+\nabla_1 z)^2  R(g)- (-1+\nabla_1 z)   R(g^0)=0.
\label{(C.17)}
\end{equation}
Consequently, the condition
\begin{equation}
R(g)=0,
\label{(C.18)}
\end{equation}
implies a simple expression of $R(g^0)$
\begin{equation}
\frac{2 \nabla_2 z}{(E G-F^2)}= (-1+\nabla_1 z)   R(g^0),
\label{(C.19)}
\end{equation}
which coincides with the Darboux eqtion (C.9) in invariant form, i.e. \cite{BI1}
\begin{equation}
\bigtriangleup_{22}z=K (1-\bigtriangleup_{1}z).
\label{(C.20)}
\end{equation} 
 
\section*{Acknowledgments}

G E is grateful to the Dipartimento di Fisica ``Ettore Pancini'' of Federico II University for
hospitality and support. D B thanks R T Jantzen for useful discussions
and the International Center for Relativistic Astrophysics Network ICRANET for
partial support. Our paper is dedicated to the memory of Stephen Hawking.

\end{document}